\documentclass{article}

\usepackage{microtype}
\usepackage{graphicx}
\usepackage{subfigure}
\usepackage{booktabs} % for professional tables

\usepackage{hyperref}

\usepackage[accepted]{icml2021}
\icmltitlerunning{Submission and Formatting Instructions for ICML 2021}

\begin{document}

\twocolumn[
\icmltitle{Hearts Gym: Learning Reinforcement Learning as a Team Event}



\icmlsetsymbol{equal}{*}

\begin{icmlauthorlist}
\icmlauthor{Jan Ebert}{jsc}
\icmlauthor{Danimir~T. Doncevic}{iek10,rwth}
\icmlauthor{Ramona Kloß}{fzj}
\icmlauthor{Stefan Kesselheim}{jsc}
\end{icmlauthorlist}

\icmlaffiliation{jsc}{Jülich Supercomputing Centre, Forschungszentrum Jülich, Jülich, Germany}
\icmlaffiliation{iek10}{Institute of Energy and Climate Research -- Energy Systems Engineering (IEK-10), Forschungszentrum J\"ulich GmbH,  52425 J\"ulich, Germany}
\icmlaffiliation{rwth}{RWTH Aachen University, 52062 Aachen, Germany}
\icmlaffiliation{fzj}{IBG-1: Biotechnology, Forschungszentrum Jülich, Jülich, Germany}

\icmlcorrespondingauthor{Stefan Kesselheim}{s.kesselheim@fz-juelich.de}
%\icmlcorrespondingauthor{Eee Pppp}{ep@eden.co.uk}

\icmlkeywords{Machine Learning, Reinforcement learning, Multi-agent, Card games, Teaching machine learning}

\vskip 0.3in
]



\printAffiliationsAndNotice{}  % leave blank if no need to mention equal contribution

\begin{abstract}
%This document provides a basic paper template and submission guidelines.
%Abstracts must be a single paragraph, ideally between 4--6 sentences long.
%Gross violations will trigger corrections at the camera-ready phase.

Amidst the COVID-19 pandemic, the authors of this paper organized a Reinforcement Learning~(RL) course for a graduate school in the field of data science. We describe the strategy and materials for creating an exciting learning experience despite the ubiquitous Zoom fatigue and evaluate the course qualitatively. The key organizational features are a focus on a competitive hands-on setting in teams, supported by a minimum of lectures providing the essential background on~RL.
The practical part of the course revolved around \emph{Hearts Gym}, an RL~environment for the card game \emph{Hearts} that we developed  as an entry-level tutorial to~RL. Participants were tasked with training agents to explore reward shaping and other RL~hyperparameters. For a final evaluation, the agents of the participants competed against each other.
\end{abstract}

\section{Introduction}\label{sec:intro}
In the last decades, many subfields of Machine Learning~(ML) have made impressive progress and drawn the attention of many scientists who are transforming their disciplines by applying and developing ML~methods. However, Reinforcement Learning~(RL), while producing notorious headlines about breakthroughs of artificial intelligence, lacks broader application to ordinary scientific problems because its implementation and adaptation require proficient practitioners. 
While teaching material and self-teaching opportunities for Supervised Learning in particular are omnipresent, opportunities to get started and collect experience with~RL in a guided setting are scarce. 

With \textit{Hearts Gym}, we created an opportunity for such hands-on experience in the form of a training course for the students of the graduate school HDS-LEE (Helmholtz School for Data Science in Life, Earth and Energy) located at Forschungszentrum Jülich (FZJ, Jülich Research Centre). %, one of Germany's largest research centres.
This course has been established during the COVID-19 pandemic, a time during which virtually all academic staff has been confined to home office, resulting in low engagement and tiredness from long hours of video conferencing, so-called \textit{Zoom fatigue} (independent of the  tool)~\cite{mcculloch_2020}. As an antidote to this Zoom fatigue, we decided to maximize the focus on hands-on experience and to further spice up the planned training course by adding the element of competition. The course concept was proposed by the HDS-LEE students, and implemented in the frame of a \emph{voucher}, a support function  offered by the AI~consultants of Helmholtz~AI, a platform fostering the use of~AI in science. 

The Hearts Gym workshop is an RL~course based on the multi-player card game Hearts. Around 40~course participants were distributed in groups and trained RL~agents that would play Hearts against agents based on manually implemented rules and other RL~agents. For evaluation, at the end of the event, the agents were put to competition in two semifinals and a final, and an overall winning team was picked. In this paper, we describe the concept of the training event, the experience, the feedback, and provide an outlook to future events. We hope that this description motivates others to use similar educational concepts. Our teaching material is available under a GPLv2 license on GitHub\footnote{\url{https://github.com/HelmholtzAI-FZJ/hearts-gym} }.

This paper is organized as follows. In Sections~\ref{sec:SotA} and~\ref{sec:hearts}, we review related RL~teaching materials and the basic rules of the game Hearts. The format of the Hearts Gym event is outlined in Section~\ref{sec:format}. Section~\ref{sec:code} provides an overview over the applied teaching material. Observations and feedback from the event are summarized in Section~\ref{sec:results}. We conclude with an outlook on future development and application of the course concept. 

\section{Related Approaches}\label{sec:SotA}
%\todo{Maybe focus this more on multi-agent stuff? Otherwise it's a lot.}
%NOTE: I(Danimir) checked 2021 and 2020 editions of the same workshop. No reinforcement learning!
A large number of online resources exist to get familiar with the theoretical background on~RL, e.g., teaching materials shared by universities or tutorials provided on the websites of ML~software packages. Here, we focus on practical materials, i.e., (libraries of) RL~environments made available to the public such that the user can focus on specifying the agent and the reward and training the agent, and sometimes specifying the environment.
Typically such environment libraries are offered by universities, big IT~companies with strong ties to AI~research, or independent projects. %Even though the application is preset by the environment, some are more education-centered while others truly place the application in center (especially simulated autonomous driving)

Most similar to Hearts Gym is \textit{RLCard}~\citep{zha2019rlcard}. RLCard provides open-source tools for RL~research in card games. The project highlights the relevance of card games for current RL~research questions, as card games constitute a challenging but ultimately intuitively understandable multi-agent, imperfect information setting. Hearts is not readily available through RLCard. 
Other RL~libraries that feature an interface for card game environments are \textit{OpenSpiel} and \textit{PettingZoo}. 
OpenSpiel~\citep{LanctotEtAl2019OpenSpiel} is a collection of RL~environments and algorithms for training agents for a plethora of games, including trademark board games. Regarding card games, some well-known games such as \textit{Bridge} are featured, but Hearts is not as of yet.
PettingZoo~\citep{terry2020pettingzoo} is another library featuring many RL~environments that focuses on environments allowing for \emph{multi-agent}~RL, including some card games. In some cases, PettingZoo acts as a multi-agent extension to \textit{Gym}.
OpenAI Gym is one of the most famous and fundamental RL~libraries, implementing a common environment API~\citep{brockman2016gymwhitepaper}. Gym provides close to a hundred RL~environments, even without counting third-party environments. Notable categories are the famous Atari games~\citep{mnih2015human}, environments built on the physics engine \textit{MuJoCo}~\citep{todorov2012mujoco}, e.g., to simulate robots, and \textit{cart pole}, a typical baseline environment and first project in RL~courses.

Other noteworthy public RL~environments from purely educational providers are a 1~vs.~1 snowball fight from a deep RL~course by HuggingFace~\citep{huggingface_snowball} and \textit{Deep Traffic}~\citep{fridman2018deeptraffic}. 
Deep Traffic appears to have a comparable scope to Hearts Gym. It provides an environment and an agent template to train car agents that navigate a busy highway road as fast as possible. Users have to tune hyperparameters, among which the most important is the neural network for Deep Q-Learning. Further, Deep Traffic features a 2D visualization of the environment and a leader board in its Git repository.

Especially for self-driving cars, RL~environments with modern computer graphics exist. For instance, \textit{AirSim}~\citep{airsim2017fsr} is a simulator for autonomous vehicles built with a game design engine. In AirSim, researchers can try out algorithms for computer vision and reinforcement learning simultaneously. Further, \textit{DeepRacer} by Amazon~\citep{balaji2019deepracer} is a cloud-based 3D race simulator that has its own global competitive leagues.
The element of competition is also central in \textit{AIArena}~\citep{sc2aiarena}, an environment and platform for competition between AI~agents in the computer game StarCraft~II. The participating agents can be trained, e.g., with DeepMind's environment \textit{PySC2}~\citep{vinyals2017starcraft}.% or even by using third-party environments (e.g.,~\citep{burnysc2,sharpysc2}).

\section{About the Game of Hearts}\label{sec:hearts}

Hearts is a simple 4-player trick-taking card game played with a regular deck of 52~cards in which the goal of a player is to \textit{avoid} collecting penalty cards. The essential rules are stated here.
Following suit is mandatory and the highest card played in the same suit as the first card of a trick wins the trick. Penalty cards include all 13~cards of hearts (1~penalty each) and the queen of spades (13~penalty!). The suit hearts can only be the leading suit once it has been played in a preceding trick in the game by a player who was not able to follow suit. However, no penalty cards, including hearts, can be played in the first trick, in which the 2~of clubs always leads. We note for the sake of completeness that Hearts has some additional complicating rules which enrich the strategic options of players and therefore offer challenges for advanced RL~agents. One of these opens up a high-risk, high-reward strategy with extremely sparse reward information for RL~agents.

The basic strategies of Hearts %as well as information that is important to keep track of 
reveal themselves intuitively to human players, providing starting points for reward shaping.
Additionally, the implementation of competitive rule-based agents that can serve as benchmarks is a rather straightforward practice.
As a further advantage, with just 13~tricks to play to finish a game, computing a full game's policy gradients, used for training, has manageable cost for personal computers.
Moreover, Hearts can be played by participants \textit{in vivo}, in preparation for and during the event, to get to know each other and to get a first feeling for the game. 
Finally, being a 4-player game, various competitions between participants can conclude the course.
For these reasons, Hearts is a great game for a hands-on RL~course.

\section{Event Format}\label{sec:format}

%Program:
To enable strong engagement and enable a positive social experience, the chosen course format is largely based on a hackathon with few introductory lectures spread across the initial phase. During four core hours in the mornings of four subsequent days, the students worked on the RL~algorithms. The afternoons were deliberately left vacant as the strong focus of real-world hackathons cannot be established in an online format. This free time was used for work on the hackathon only by a subset of the participants. The participants of the event were graduate students with a focus in applied machine learning. Therefore, it was possible to assume a decent background in mathematics and general~ML.
%Input Talk(s):
The aim behind the selected teaching material was to convey basic concepts of~RL, such as what environments, actions, rewards, and the ideas behind RL~algorithms are, e.g., maximizing cumulative future reward. It is fair to say that the priority of the course was to provide a hands-on and social experience rather than in-depth understanding of the applied RL~methods.

To this end, the participants watched a 15-minute video \emph{An introduction to Reinforcement Learning} %from the channel \emph{Arxiv Insights}
as a quickstart~\cite{arxivinsights}. This knowledge was deepened in a reading phase of the section \emph{Part~1: Key Concepts in~RL} of OpenAI's deep~RL introduction \emph{Spinning Up}~\cite{spinningup}. Furthermore, the participants were referred to the freely available, standard RL~textbook of Sutton and Barto~\cite{sutton2018reinforcement}, two UCL~courses on~RL~\cite{silver2015, hasselt2018}, and two sections with practical RL~tips and additional resources of the online documentation of the software package \emph{Stable Baselines3}~\cite{stable-baselines3}. This selection, in combination with the ready-to-go software environment, enabled students to get started quickly, without over-simplifying the field. 

%\todo{Es gab eine Einführung zu Hearts und Code Safari; Hearts-Verständnis wurde durch praktisches Spiel zwischen den Teilnehmern verstärkt. Sozialer Aspekt davon, eventuell Website Linken.}
In addition to the self-study material, we demonstrated the source code of our software environment~(explained in more detail in Section~\ref{sec:code}) and its structure in a \emph{code safari}. To get familiar with the card game, the rules were explained to the participants. As an ice breaker and to reinforce understanding, participants were encouraged to start their group work by playing an online Hearts game. 
We scheduled the remainder of the available time for preparing the Hearts-playing agents in a team setting. 
To this end, the participants teamed up in eight groups of 4--6~people. The groups were formed before the workshop started and were composed of at least one PhD student with advanced experience in~ML and one who is proficient in Python. %The groups were either autonomously assembled in advance or drawn by lot according to the above criteria.

The two mentors of the course were available via a group chat set up for the hackathon. In addition, the mentors occasionally checked into each group's video conferencing room. This provided the mentors with a picture of how the hackathon is going while not overwhelming participants, but also enabled support for groups that were less inclined to ask for help on their own.
Following the hackathon convention, participants could and did use the time between the scheduled sessions for further work on their agents.
Submissions of agents were due during the fourth and final session. Each team could submit one RL~agent and one rule-based agent, which were pitted against agents of other teams in respective tournaments with a knockout format. In each tournament round, 4~agents played several matches of Hearts against each other and the two best-performing agents advanced to the next round. Overall winners in the~RL and rule-based categories were determined in the final.

\section{Learning Environment Implementation}\label{sec:code}

%% integrate me
The learning environment was constructed following the principle that simplicity and ease of use are clearly favored over complexity and performance. Therefore, we chose a Python implementation using widespread RL~libraries and minimal further dependencies. 
To get started on the practical side of the course, participants only required a~PC, Git, and a Python distribution. Then, Hearts Gym could be retrieved from Git and its dependencies automatically downloaded.
%% end

Hearts Gym was implemented using the multi-agent environments of RLlib~\citep{liang2018RLlib}. % It is part of the Ray suite maintained by UC~Berkeley. 
The concept of RL~frameworks entails that users provide implementation of the key elements of~RL, such as environment, policy, etc. The framework defines the relevant interfaces between these modules, in order to obtain a composable architecture.  Dependencies were deliberately kept minimal in order to have an easily maintainable environment with the least amount of possible failure modes. To prevent technical issues during the event, two test runs with volunteer participants were done in preparation. These helped sort out portability errors and resulted in first feedback and according improvements.

%\todo{Most important software modules}: \texttt{HeartsGame}. \texttt{RandomPolicy} \texttt{ObservedGame} The wrappers.

The learning environment thus provided the required implementations for a wide range of possible policies. Hence, the participants were able to explore a lot of different strategies simply by adjusting a configuration file. This also allows participants with a weaker background or strong limitations regarding the available time to take full participation in the activity. With intuitive alterations of the configuration file, participants were already able to make successful improvements to the agents. More advanced learners could dive deeper into the code and significantly change the training procedures, implement new rule-based policies, or make even more complex alterations. One complex modification would be the implementation of new algorithms. This could be encompassed as a learning exercise in a more advanced version of the hackathon.

The most important and fundamental code modification was the implementation of a custom reward function. This reflects the fact that \emph{reward shaping} is a key ingredient for successful~RL. 
In addition, participants were encouraged to implement their own rule-based agents. This was possible with relative ease as we also implemented a simple deterministic policy which could serve as a skeleton for participant implementations. 

For the remote evaluation, a client-server structure was created. This also allowed teams to set up their own servers and evaluate their agents against one another. One challenge was the collection of results in a relatively short time frame, while still allowing for internet connections with some latency. To this end, on the one hand, we implemented an auto-kick-system that replaces players who take too long to respond with randomly acting bots. On the other hand, many games are played and communicated simultaneously.
The servers stayed active for the duration of the course. Some teams used the servers as ``sparring grounds'' to coordinate evaluations of their agents against each other even during the hackathon, which improved communication between groups. At the end of the event, the servers were used as an arena for the finals, for which evaluation was carried out live in a video conference.

\section{Event Evaluation}\label{sec:results}

One of the authors participated in the course and the hackathon and we interviewed four additional participants from two groups. Here, we qualitatively summarize their feedback. This qualitative feedback closely matches the quantitative feedback collected from 17 participants.

%OVERALL FEEDBACK
All of the interviewed participants found the interactive group setting of the course appropriate, given the pandemic-induced home office situation. Even though the online format allowed minor distractions from the hackathon, the engagement level was high.
The competitive element of the course contributed positively to this engagement level. However, the limited time frame led some participants to disregard the competitive element.
%learning high-level theory
Participants highly approved of the learning experience provided by the Hearts Gym hackathon with regard to high-level understanding of~RL. Studying the materials and participating in the hackathon, participants believed that they learned enough about~RL to competently discuss ideas of application of~RL to their field.
In addition, some participants consider the experience gathered sufficient to plan or even conduct an RL~project on their own. At least, the event has revealed existing gaps in the knowledge of individuals.
%learning (curve) of applying/using RL~environments
The setup of Hearts Gym worked for all users. However,
participants felt that proficiency in Python was a requirement to navigate the code base in the limited time. Further, differing expertise in Python tended to decrease collaboration within the groups.
In consequence, the practical learning success during the hackathon varied from group to group.
One of the teams invested more than the allotted time of four half days on the practical task and stated a noticeable day-to-day progress in handling the tools available in Hearts Gym as a result of their effort.
Other participants experienced flatter learning curves due to long feedback loops of RL~training and a high amount of options that could be manipulated. On top of that, inefficient group dynamics including screen-sharing of code and resulting pair programming slowed down progress.

%EXECUTION OF THE EVENT
%technical/HeartsGym
%The setup of Hearts Gym worked for all users but some connectivity issues with the evaluation server were reported (due to also tunneling via VPN?)
%strategy/hackathon
As a strategic approach to the challenge posed by Hearts Gym, the interviewed participants reported initially focusing on the logic behind clever play in Hearts. Experimentation with rule-based agents was common. In fact, the best agent developed during the hackathon was a sophisticated rule-based agent.
Later during the hackathon, most teams focused on reward shaping and hyperparameter tuning for~RL in a very exploratory fashion.
%tactics/reward shaping
With the default evaluation setting being play versus randomly acting agents, we observed a tendency among participants to target small tactical improvements for learning that would lead to better-than-random play. For instance, if an agent learned to drop isolated high cards in the first trick without penalty, they would typically perform better than average even with random play in the remaining tricks.
%\footnote{Recall that no penalties can be played in the first trick. Therefore, it is ``safe'' to win the trick with a high card, which helps to strengthen the remaining hand.}
% Initially, many participants targeted such low-hanging fruits, even if only to validate whether learning those rules of thumb indeed leads to better play.
Once the submission deadline approached, exploitation in the form of longer training times or larger policy networks for promising configurations was conducted by all teams.
Some groups have used high-performance computing~(HPC) resources to reduce the training times of agents and reported that the faster feedback enabled a more stringent exploration.
None of the teams spent time on analyzing the individual games played by their agents in more depth. Therefore, reward shaping was exclusively driven by trial and error based on the intuition of the participants.
In the final evaluation, all teams succeeded to submit agents that played distinctively better than random and a deserved winner could be crowned among them.

\section{Conclusion and Outlook}

Based on the experience and the feedback, the event can be considered a success. With moderate time effort, the course participants became familiar with the key concepts of~RL and had a group experience that was as immersive as the COVID pandemic allowed for.

With small tuning, the event concept can be adjusted for in-presence hackathon-like events. The duration can be altered to 2--3~full days. In-presence activities allow for much better team-building activities to strengthen participant interactions. In addition, scientific interaction, such as a poster session, have proven to be positive in other events organized by the authors. Especially in-person, it is important to speed up the feedback cycle. Providing access to HPC~center resources with many-core~GPUs is then a must. We hope that the release of our material helps other teams to set up such activities. Interested persons are encouraged to contact us. 

Another exciting aspect would be integrating the challenge into a challenges platform such as the Helmholtz Data Challenges\footnote{\url{https://helmholtz-data-challenges.de}}. Here, we imagine that agents submitted as code to such a challenges platform would be evaluated automatically with new submissions. This convenient long-time evaluation environment could create a setting of continuous progress just as other data challenges have provided.

\section{CRediT Author Statement}
This section states the authors main roles following Contributor Role Taxonomy CRediT~\cite{allen2019can}. \\
\textbf{Jan Ebert}: Conceptualization, Validation, Methodology, Software, Resources,
\textbf{Danimir Doncevic}: Writing -- Original Draft,  Conceptualization,  Evaluation, Literature Review,
\textbf{Ramona Kloß}: Coordination, Administration,
\textbf{Stefan Kesselheim}: Conceptualization, Supervision

% \section{credit author statement}
% 
% jan ebert: software development, literature review, event planning and execution, user support, writing~-- review \& editing,
% % jan ebert: methodology, software, resources,
% danimir doncevic: writing -- original draft, writing -- review \& editing, conceptualization,  evaluation, literature review,
% ramona kloß: coordination, administration,
% Stefan Kesselheim: Conceptualization, Supervision, Writing -- Review \& Editing.

\section*{Acknowledgements}

This work was partially performed as part of the Helmholtz School for Data Science in Life, Earth and Energy (HDS-LEE) and received funding from the Helmholtz Association of German Research Centres.

% Acknowledge trial run participants?
We would like to thank Ji Gao, Daniel Mayfrank, Felix Rauh, and Luisa Schuhmacher for participating in trial runs ahead of the event. Their feedback was extremely valuable for making the event run more smoothly.


\bibliography{main}
\bibliographystyle{icml2021}

%%%%%%%%%%%%%%%%%%%%%%%%%%%%%%%%%%%%%%%%%%%%%%%%%%%%%%%%%%%%%%%%%%%%%%%%%%%%%%%
%%%%%%%%%%%%%%%%%%%%%%%%%%%%%%%%%%%%%%%%%%%%%%%%%%%%%%%%%%%%%%%%%%%%%%%%%%%%%%%

\end{document}